\pdfoutput=1

\documentclass[10pt]{article}

\usepackage{tcolorbox}
\usepackage{xspace}

\usepackage{ifthen}
\newboolean{showcomments}
\setboolean{showcomments}{true} 

\ifthenelse{\boolean{showcomments}}{
	\newcommand{\nbc}[3]{
		{\colorbox{#3}{\bfseries\sffamily\scriptsize\textcolor{white}{#1}}}
		{\textcolor{#3}{\sf\small$\langle$\textit{#2}$\rangle$}}}
	
}{
	\newcommand{\nbc}[3]{}

}

\newtcolorbox{resultsbox}[1][]
{
colframe=gray!100, 
colback=white!100, 
coltitle=white,
title=#1 
}

\newenvironment{results}[1][]{
    \begin{resultsbox}[#1]
 }{
   \end{resultsbox}
}

\newcommand{\evo}{{\sc EvoMaster}\xspace}
\newcommand{\uncermaster}{{UncerMaster}\xspace}
\newcommand{\uncersense}{{UncerSense}\xspace}
\newcommand{\uncermeter}{{UncerMeter}\xspace}
\newcommand{\uncerfuzz}{{UncerFuzz}\xspace}
\newcommand{\uncermml}{{UncerMML}\xspace}

\newcommand{\checkout}{\emph{Checkout}\xspace}
\newcommand{\payment}{\emph{Payment}\xspace}
\newcommand{\fraud}{\emph{Fraud}\xspace}
\newcommand{\order}{\emph{Order Confirmation}\xspace}
\newcommand{\shipping}{\emph{Shipping}\xspace}

\newif\ifrevision

\ifrevision
  \newcommand{\revision}[1]{{\color{blue}#1}}
\else
  \newcommand{\revision}[1]{#1}
\fi

\usepackage{authblk}                
\usepackage[square,numbers]{natbib} 
\usepackage{amsmath}                
\usepackage{amssymb}                
\usepackage{hyperref}
\usepackage{booktabs}
\usepackage{geometry}
\title{Fuzzing Microservices in Face of Intrinsic Uncertainties
}

\author[1]{Man Zhang}
\author[1]{Tao Yue\thanks{Corresponding author}}
\author[2]{Andrea Arcuri}

\affil[1]{Beihang University, China}
\affil[2]{Kristiania University College and Oslo Metropolitan University, Norway}

\affil[ ]{\{manzhang, yuetao\}@buaa.edu.cn, andrea.arcuri@kristiania.no}
\date{}

\begin{document}

\maketitle

\begin{abstract}
The widespread adoption of microservices has fundamentally transformed how modern software systems are designed, deployed, operated and maintained. However, well-known microservice properties (e.g., dynamic scalability and decentralized control) introduce inherent and multi-dimensional uncertainties. These uncertainties span across inter-service interactions, runtime environments, and internal service logic, which manifest as nondeterministic behaviors, performance fluctuations, and unpredictable fault propagation. Existing approaches do not have sufficient support in capturing such uncertainties and their propagation in industrial microservice systems, and these approaches mostly focus on single-service testing. In this paper, we argue for a novel paradigm: ``uncertainty-driven'' and ``system-level'' microservice testing. We outline key research challenges, including the modeling and injection of uncertainties and their propagation, causal inference for fault localization, and multi-dimensional analyses and assessment of uncertainties and their impact on system quality. We propose an architecture for continuous uncertainty-driven and system-level microservice fuzzing, which integrates service virtualization, uncertainty simulation, adaptive test generation and optimization\revision{, and illustrate it with an e-commerce example we developed}. Our goal is to inspire the development of scalable and automated system-level testing methods that improve the dependability and resilience of industrial microservice systems, with the explicit consideration of uncertainties and their propagation. 
\end{abstract}


\section{Introduction}

Microservices architecture has emerged as a foundational model for building advanced technologies (e.g., cloud and edge computing) and modern applications (e.g., industrial services), driven by its inherent technical strengths, such as service decoupling (enabling independent deployment), fault isolation (preventing cascading failures), elastic scalability (adapting to dynamic workloads), and polyglot persistence (supporting heterogeneous data stores)~\cite{newman2021building,richards2025fundamentals}. 
For example, the microservices architecture of Alibaba (\revision{a Fortune 500 corporation})
utilizes a modular microservices approach to enable the dynamic integration of multi-source data and to support real-time decision-making in areas like personalized pricing and fraud detection~\cite{luo2022depth}. 
Similarly, MindSphere of Siemens (\revision{a Fortune 500 corporation})
is an industrial IoT platform that leverages microservices to orchestrate heterogeneous edge devices and cloud-based analytics, ensuring fault-tolerant data processing in mission-critical manufacturing workflows~\cite{lechowski2022emerging}.
Meituan (\revision{a Fortune 500 corporation}) developed online and on-demand delivery platforms with microservices, offering their services to more than 600 million users~\cite{zhang2023rpc,zhang2024seeding,zhang2025industry}.
According to recent analytics, Edge Delta estimates that 94\% of companies worldwide rely on cloud computing, with related expenditures reaching into the hundreds of billions of dollars~\cite{cloudcomputing2024}.

As the adoption of microservices continues to grow across industries, ensuring the dependability of microservice systems has become critical. Industrial microservices are often characterized by complex business logic, further compounded by dynamic scalability, heterogeneous architectures, and distributed topologies. These characteristics introduce multi-dimensional uncertainties that affect both inter-service interactions (e.g., chains of API calls, message queues) and internal service implementations (e.g., non-deterministic algorithms, resource contention). Such uncertainties can lead to service anomalies (e.g., timeout cascades), performance degradation (e.g., latency spikes), and even cascading failures (e.g., database connection pool exhaustion), significantly jeopardizing system's dependability. 

Therefore, it is essential to establish a comprehensive testing framework that can explicitly incorporate various types of uncertainties inherent in microservice architectures and their operating environments. 
By explicitly considering these uncertainties and their propagation during testing, the testing framework can simulate real-world conditions and stress-test the system, which enables organizations to proactively identify potential points of failures, mitigate risks, and optimize system resilience. 
Otherwise, critical issues such as service downtime, slowdowns, or cascading failures might go undetected, leading to compromised quality and unreliable service delivery in operating environments. 
\revision{In this paper, we propose a conceptual architecture that defines the required components and key workflows to facilitate continuous uncertainty-driven and system-level microservice fuzzing.
The proposed framework has not yet been fully implemented and it is intended to serve as a foundation for future development and empirical evaluation.
}

Structure: Section~\ref{challenges} discusses the key challenges to realize uncertainty-driven microservice testing; Section~\ref{literature} discusses the state of art of microservices testing and uncertainty-aware software engineering, followed by our vision towards an  uncertainty-driven microservice testing (Section~\ref{future}); \revision{We illustrate our framework with an e-commerce example in Section~\ref{sec:example} and provide discussions in Section~\ref{sec:discussions}.} We conclude the paper in Section~\ref{conclusion}.

\section{Challenges in Uncertainty-driven Microservice Fuzzing}\label{challenges}
The literature on microservice testing~\cite{golmohammadi2023testing}
predominantly focuses on deterministic scenarios (e.g., single-service API fuzzing) and lacks systematic ways to guide industrial practices in mitigating hidden risks arising from complex interactions, and inadequately addresses uncertainties in microservices (e.g., non-deterministic inputs, dynamic propagation paths). The demand for revolutionizing microservice testing has dramatically increased,  especially when the widespread adoption of microservices in mission-critical domains such as financial transaction platforms (e.g., AliPay and Apple Pay) and real-time healthcare systems (e.g., 
patient monitoring systems in the Intensive Care Units).

In the rest of this section, we detail three challenges faced by the existing approaches in microservice testing, focusing on multi-dimensional uncertainty sources (Section~\ref{sub:multi-dimentional}), dynamic uncertainty propagation (Section~\ref{sub:dynamic}), and combinatorial explosion of microservice uncertainties (Section~\ref{sub:combinatorial}). \revision{In Section~\ref{subsec:deltaWithDistributedSystems}, we also differentiate uncertainties in microservices with those in traditional distributed systems. }

\begin{figure}[t]
    \centering
    \includegraphics[width=.9\linewidth]{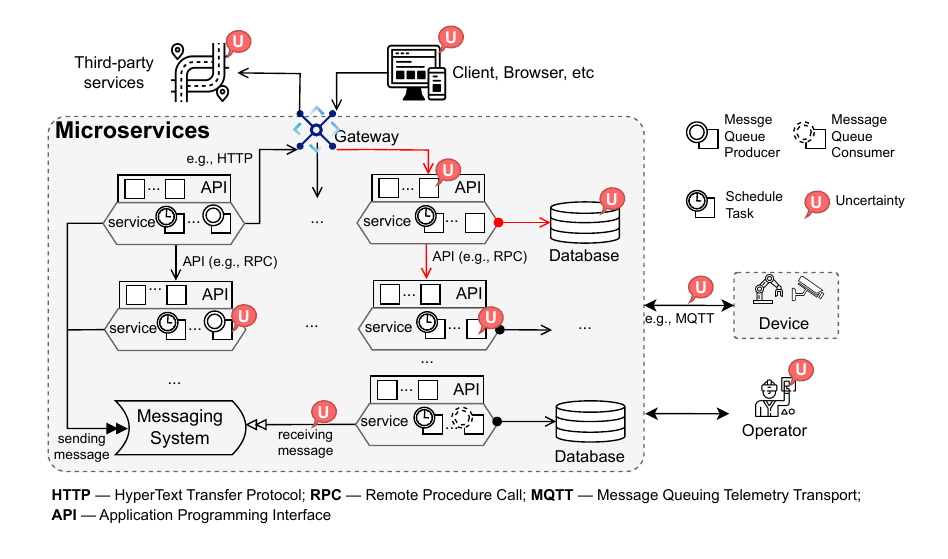}
    \caption{Illustrating Potential Occurrences of Uncertainties in Microservices. Uncertainties may occur in third-party services, client sides, APIs, database, message queuing, scheduling tasks, etc.}
 \label{fig:challenges in microservices testing}
\end{figure}

\subsection{Multi-dimensional microservice uncertainties}
\label{sub:multi-dimentional}
The architecture of microservices is inherently distributed, hence microservices face multi-dimensional uncertainties and complex interactions among them. As illustrated in Figure~\ref{fig:challenges in microservices testing}, in industrial application contexts, such uncertainties may originate from network fluctuations (e.g., latency, jitter), heterogeneous environments (e.g., resource contention), service dependency anomalies (e.g., version compatibility issues), and data flow exceptions (e.g., message queue accumulation).

There are also other uncertainties such as uncontrollable external inputs (e.g., abnormal user request parameters), inherent randomness introduced by employing Artificial Intelligence (AI) technologies (e.g., prediction fluctuations in machine learning models), and system-internal stochastic function calls (e.g., random load balancing algorithms). These uncertainties exhibit not only diverse types, but also dynamic and concealed characteristics. For instance, transient resource contention at an edge node might only be triggered under specific loads, while version upgrades of dependent services might implicitly introduce protocol parsing errors.

Furthermore, uncertainties exhibit deep coupling across hierarchical levels (ranging from hardware to systems and applications) and cross-service scenarios such as interactions and dependencies between upstream and downstream services in call chains. For example, malicious user inputs might trigger AI model misjudgments, while random load balancing strategies could amplify network jitter effects, potentially triggering cascading failures.
Existing approaches that rely on static topology modeling or limited-dimensional monitoring (e.g., analyzing only logs and network metrics) often lack capabilities of dynamically understanding spatial-temporal correlations of multi-dimensional uncertainties. Additionally, these approaches fail to quantify hidden risks from inherent uncertainties (e.g., nondeterministic programs), which leaves test input generation without sufficient guidance. 

Therefore, unified modeling and dynamic detection of multi-dimensional uncertainties form the foundational prerequisite to construct uncertainty-oriented fuzzing frameworks and detect complex failure scenarios.

\subsection{Dynamic uncertainty propagation}
\label{sub:dynamic}

Inter-dependencies among microservices allow uncertainties localized to one service node (e.g., a single service timeout) to rapidly propagate through call chains, message queues, or shared storage, which consequently trigger cascading failures or even system-wide avalanches. For instance, a response delay in an order service 
might generate a high volume of retry traffic to
the payment service, subsequently exhausting database connection pools and potentially paralyzing the entire transaction workflow. Such uncertainty propagation exhibits nonlinear, asynchronous, and path-indeterminate characteristics. Specifically, propagation paths may dynamically change due to load-balancing strategies, states of circuit-breakers (for isolating faulty services), or message queue backlogs. Therefore, it is very difficult to rely on typical metrics (e.g., error rates) to quantify the intensity and boundaries of uncertainty propagation. More critically, observability limitations in distributed systems (e.g., incomplete trace data) can obscure critical propagation nodes, resulting in scenarios where localized anomalies are visible, but impacts remain opaque.

Existing approaches that rely on static dependency graphs or simple error-rate statistics fail to model the causal logic (e.g., triggering conditions for cascading failures) and its implications (e.g., robustness degradation) of uncertainty propagation. Metrics such as Mean Time To Recovery (MTTR) further fall short in quantifying dynamic threats to system resilience (e.g., hidden bottleneck services). 

Therefore, we call for research on studying causal inference and assessment of uncertainty propagation of microservices, tracking and quantifying uncertainty propagation and its impact on system quality properties (e.g., availability). Doing so is essential for prioritizing high-risk paths and underpin uncertainty-aware fuzzing methodologies.

\subsection{Combinatorial explosion of microservice uncertainties}
\label{sub:combinatorial}
Industrial microservices systems often require hundreds to thousands of functional services to support complex business logic. For example,  the shopping platform developed by Meituan (\revision{a Fortune 500 corporation})
comprises over 2,000 services~\cite{zhang2023white}, and WeChat (the flagship mobile super-app of Tencent, \revision{a Fortune 500 corporation})
includes more than 3,000 services as of 2018~\cite{zhou2018overload}. These services form a high-dimensional dependency topology: a single business request processing typically triggers multi-level, cross-service and distributed call chains (e.g., rendering a single web page of Amazon, \revision{a Fortune 500 corporation}, 
invokes approximately 100–150 APIs~\cite{zhou2018fault}). Consequently, the combinatorial space of uncertainties grows exponentially with the number of services. For instance, 3,000 services could yield $2^{3000}$ abnormal state combinations. Existing testing methods 
are inefficient at generating high-quality test cases, facing a needle-in-a-haystack dilemma: excessive resources are wasted on low-risk combinations (e.g., isolated anomalies in non-critical services), while high-risk scenarios (e.g., concurrent failures in multiple strongly dependent services within payment workflows) are uncovered.

Existing testing techniques are unable to intelligently balance propagation risks of cross-service combinations of uncertainties, business priorities, and critical scenarios, leading to poor testing efficiency and fault detection rates. Therefore, developing adaptive testing strategies and coverage optimization mechanisms that account for combinations of microservices uncertainties is critical to reconcile the tradeoff between constrained testing resources and combinatorial explosion. This is the pragmatic objective of uncertainty-aware fuzzing research for industrial microservices systems.

\revision{

\subsection{Uncertainty in Microservices and Traditional Distributed Systems}\label{subsec:deltaWithDistributedSystems}

Many uncertainties we discussed above are not unique to microservices but are shared across general distributed systems.
Examples such as network latency variability, transient dependency failures, and data inconsistency due to asynchronous communication are well-known challenges in traditional distributed computing paradigms, including Service-Oriented Architecture (SOA) and client-server systems~\cite{laskey2009service,van2017distributed}. 
Network latency and partial failures are fundamental concerns in distributed systems that drive adoption of failure detectors and consistency protocols. 
Transient dependency failures and cascading effects have also been studied extensively in the fault tolerance literature, where techniques such as timeouts, retries, and idempotency are used to mitigate their impact~\cite{cristian1991understanding,xing2020cascading}.

However, due to architectural characteristics, microservices further amplify these uncertainties.
For instance, traditional distributed systems~\cite{dragoni2017microservices}.
Moreover, each service can scale and evolve independently, which leads to a larger state space of possible configurations and interactions where uncertainties can propagate in non-trivial ways~\cite{lewis2014microservices}.

Recent industrial systems further integrate AI- and LLM-based components into microservice architectures~\cite{al2022ai}, which introduces new forms of uncertainties by interacting with already existing challenges intrinsic to distributed systems.
For example, inference latency, model nondeterminism, probabilistic outputs, and model version drift can introduce additional variability in service responses. 
Specifically, LLM-based services are sensitive to input phrasing, prompt changes, and contextual dependencies. All these may lead to uncertain system behaviors that are difficult to validate. 
Furthermore, after being embedded in microservice workflows, their stochastic and evolving nature can amplify uncertainty propagation across services, and hence complicate fault localization, and eventually affect the end-to-end reliability of the systems.

Furthermore, AI- and LLM-driven decision-making often relies on external model-serving platforms and large-scale inference infrastructures, which introduces additional dependencies and potential failure modes. 
For instance, resource contention in GPU clusters, dynamic model updates, and opaque performance characteristics may increase latency fluctuations and cascade failures in microservice systems. 
Consequently, uncertainty in modern microservices increasingly arises not only from traditional distributed-system factors but also from the inherent variability of learning-based components.

Although many uncertainties are shared with distributed systems in general, along with the advance of AI- and LLM-based techniques, modern microservices introduce higher degrees of variability, emergent cross-service behaviors, and operational complexity that make uncertainty quantification and testing particularly challenging. 
}

\section{The State of Art}\label{literature}
Microservices testing and uncertainty-aware software engineering in complex software systems (e.g., Cyber-Physical Systems (CPS)) has gained significant attention from both academia and industry in the last decade. In this section, we discuss the state-of-the-art from these aspects: microservices fuzz testing, uncertainty classification and characterization, uncertainty quantification, and uncertainty-aware testing.  

\subsection{Microservices testing}

\subsubsection{API fuzzing}
Most existing efforts in microservice testing focus on single-service testing, particularly API fuzzing. Currently, mainstream API paradigms for microservices include REST, RPC, and GraphQL. In the rest of the section, we discuss the literature of testing each type of APIs. 

REST defines API design guidelines for resource access and manipulation using standard HTTP methods, making it suitable for open platforms and third-party integrations. Its broad adoption has spawned a vast ecosystem of open-source tools, and REST API fuzzing has consequently become the most extensively studied area. Golmohammadi et al.~\cite{golmohammadi2023testing} surveyed 92 articles on REST API testing published from 2009 to 2022, and revealed that 72\% of the proposed methods belong to black-box testing. 
Most \textbf{REST API black-box testing} approaches leverage API specifications (e.g., OpenAPI Specification, OAS, formerly known as Swagger) to parse accessible requests and validate interface behaviors using runtime request/response data. For instance, Restler~\cite{godefroid2020intelligent, atlidakis2019restler, atlidakis2020checking, barlas2022exploiting, godefroid2020differential} infers API request dependencies from OAS and runtime data to construct test sequences, effectively probing REST API behaviors. Morest~\cite{liu2022morest} formalizes REST API dependencies as property graphs to capture relationships between operations and data structures, enabling test case generation and fault detection. RestCT~\cite{wu2022combinatorial} is a combinatorial testing method that analyzes request/parameter dependencies and constraints via OAS and runtime data, generating test cases to exhaustively cover potential API input combinations. 
Kim et al.~\cite{kim2023adaptive, kim2023enhancing} introduced reinforcement learning and NLP-driven techniques to infer API logic and constraints for test generation. 
Schemathesis~\cite{hatfield2022deriving} is a property-based testing tool that derives test oracles from OAS semantic rules to validate REST API compliance. 
Rueda et al.~\cite{segura2017metamorphic} and Chen et al.~\cite{chen2018metamorphic} leveraged REST design principles (e.g., idempotency of HTTP methods like GET) to define metamorphic relations for effective test validation.
DeepREST~\cite{corradini2024deeprest} overcomes OAS limitations by using curiosity-driven reinforcement learning to uncover hidden API logic and constraints and hence guide test generation.
Unlike OAS-based black-box testing methods, \textbf{REST API white-box testing} approaches delve into internal system implementations, and combine source code analysis, dynamic execution traces, and metrics such as code coverage, taint analysis, and SQL query interception. For instance, Pythia~\cite{atlidakis2020pythia} is an approach that extends Restler with code coverage guidance but requires manual pre-configuration for instrumentation. \evo~\cite{zhang2021adaptive,arcuri2019restful,zhang2024seeding,zhang2023rpc,arcuri2024advanced,arcuri2020testability} fully supports automated white-box fuzzing for APIs. LT-MOSA~\cite{stallenberg2021improving} employs hierarchical clustering of API call sequences to optimize test case generation.

GraphQL was designed to address limitations of REST APIs in data access. For example, GraphQL allows clients to precisely specify what data to retrieve from a graph-like structure~\cite{hartig2018semantics, taelman2018graphql}, which resolves inherent issues in REST such as over-fetching and under-fetching. Similar to REST, GraphQL provides API specifications (e.g., GraphQL Schema) to parse accessible requests, enabling black-box testing without source code access.
A recent literature review conducted by~\cite{quina2023graphql} highlights the scarcity of GraphQL automated testing research, as the authors managed to identify only five black-box methods~\cite{hatfield2022deriving, karlsson2021automatic, vargas2018deviation, zetterlund2022harvesting} and one white-box approach~\cite{belhadi2022evolutionary, belhadi2024random} built on top of \evo. 
For instance, Hatfield-Dodds and Dygalo~\cite{hatfield2022deriving} proposed to derive structural and semantic rules from GraphQL schema using the Hypothesis framework~\cite{loscher2018automating} to generate test cases and oracles. Karlsson and Causevic~\cite{karlsson2021automatic} proposed to extract data types and relationships from schema, employ randomized strategies to generate data or user-defined data to send requests, and then validate responses via HTTP status codes and schema compliance. Zetterlun et al.~\cite{zetterlund2022harvesting} proposed AutoGraphQL, a tool leveraging real-world GraphQL queries from operations to guide test case generation and detect regressions. Vargas et al.~\cite{vargas2018deviation} proposed four divergence rules to mutate existing test cases, and then compare the original and variant outcomes to assess API robustness.

Remote Procedure Call (RPC) is a communication protocol and mechanism that enables cross-service, cross-process function or procedure invocations~\cite{birrell1984implementing}. By abstracting the complexity of underlying communication, RPC makes remote service calls as straightforward as local function calls. Furthermore, RPC frameworks typically employ high-efficiency serialization and deserialization and lightweight communication protocols to minimize network payloads and optimize transmission performance. These features have made RPC a cornerstone for enterprise-scale microservice systems~\cite{newman2021building}.
Unlike REST and GraphQL, RPC lacks a unified API specification. Instead, RPC opts for flexible interface definitions. For instance, RPC frameworks, such as
Thrift,\footnote{\url{https://thrift.apache.org/docs/idl}} gRPC\footnote{\url{https://protobuf.dev/overview/}} and Tars,\footnote{\url{https://tarscloud.github.io/TarsDocs_en/base/tars-protocol.html}} have their own interface definition languages (IDL) to define service interfaces, and support the generation of code (e.g., client stub) in multiple programming languages. 
However, IDL-based API specifications are not a mandatory requirement for RPC implementation. For example, frameworks such as Dubbo,\footnote{\url{https://dubbo.apache.org/en/}} SOFARPC,\footnote{\url{https://www.sofastack.tech/en/}} and Mota\footnote{\url{https://github.com/weibocom/motan}} support defining and exposing interfaces directly through Java interfaces, configured via framework-specific settings. RPC API testing approaches are currently limited. Notable efforts include the RPC API testing method based on \evo~\cite{zhang2023white, zhang2024seeding}, and Zero-Config proposed by Wang et al.~\cite{wang2023zero}. Zero-Config analyzes Protobuf-based API specifications and uses LibFuzzer to automatically generate test cases for gRPC APIs developed in C++. However, the tool is currently not publicly accessible. 

\subsubsection{System-level testing of microservices}
System-level testing for microservices is relatively scarce, with works mainly focusing on regression testing based on historical data~\cite{li2024application, gazzola2023exvivomicrotest, liu2022record}, reliability testing~\cite{camilli2022microservices}, and black-box testing based on business logic or formal expressions~\cite{hillah2017automation, luo2020verification, quenum2018towards}. 
Specifically, Liu et al.~\cite{liu2022record} proposed to automatically generate mocking points by identifying the underlying system state, internal state, and external state of microservice dependencies, aiming to address the issue of online traffic playback failures caused by microservices’ dependency on state. 
Di et al.~\cite{di2024microfuzz} proposed MicroFuzz to improve code coverage and anomaly detection by using techniques such as dynamic priority seed generation, parallelized multi-stage pipelines, and cross-service context dependency tracking.
Gazzola et al.~\cite{gazzola2023exvivomicrotest} proposed ExVivoMicroTest to generate regression test cases by monitoring and recording runtime data from microservices. Chen et al.~\cite{chen2023microservice} proposed optimization methods for selecting regression test cases. Camilli et al.~\cite{camilli2022microservices} introduced MIPaRT to enable performance and reliability testing of microservices. MIPaRT generates test cases based on real operation data, calculates performance and reliability metrics by monitoring and collecting runtime data from executing test cases, and provides visualizations of collected data to support the continuous evaluation of microservices systems.
Hillah et al.~\cite{hillah2017automation} proposed a model-based testing approach to generate test cases from pre-defined models (e.g., test configuration model, service interface model, and service behavior model). Quenum and Aknine~\cite{quenum2018towards} proposed to generate test cases for microservices based on formal specification statements. To apply metamorphic testing in microservice applications, Luo et al.~\cite{luo2020verification} designed metamorphic relations based on the business logic of each application.
Almutawa et al.~\cite{almutawa2024towards} explored the use of Large Language Models (LLMs) to assist engineers in performing end-to-end microservice testing.

\begin{results}[Summary]
The literature primarily focuses on testing individual services (i.e., API fuzzing), while system-level testing methods still suffer from a heavy reliance on manual intervention. In industrial practice, system-level testing, integration testing, and end-to-end testing (which begins at the user interfaces and spans the microservices architecture and backend components such as databases to validate complete business processes) also face challenges due to insufficient automation. 
\end{results}

\subsection{Uncertainty characterization, quantification, and uncertainty-aware testing and optimization}

In this section, we discuss studies of uncertainties in software engineering from three aspects: uncertainty classification and characterization, uncertainty quantification and uncertainty-aware testing. 

\subsubsection{Uncertainty classification and characterization}
In the early stages of uncertainty modeling research in the field of software engineering, due to its popularity, the Unified Modeling Language (UML) was chosen as the primary modeling language for specifying uncertainties. For example, Ma et al.~\cite{ma2011fuzzy} proposed a fuzzy UML data modeling approach, which is translated into fuzzy description logic for the correctness verification of fuzzy attributes. Riebisch et al.~\cite{riebisch2002uml} proposed to model uncertainty in UML use case diagrams, sequence diagrams, and state machines. Motameni et al.~\cite{motameni2010transforming} proposed to transform UML state machines into fuzzy Petri nets. To support test generation, Garousi et al.~\cite{garousi2008traffic} modeled temporal uncertainty within UML sequence diagrams. However, these early approaches are preliminary and insufficient in effectively handling complex and multi-dimensional uncertainties.

Subsequently, Zhang et al.~\cite{zhang2016understanding} proposed U-Model, which defines uncertainty and its related concepts from a software engineering perspective and clarifies their relationships at the conceptual level. Building on this, a more comprehensive UML-based uncertainty modeling language, named UncerTum~\cite{zhang2019uncertainty}, was proposed. 
Later, the international standard Precise Semantics for Uncertainty Modeling (PSUM) 1.0 Beta was adopted at OMG\footnote{Precise Semantics for Uncertainty Modeling (PSUM): https://www.omg.org/spec/PSUM/1.0/Beta1/About-PSUM}, which unifies the understanding of uncertainty in the software engineering domain and provides a standardized and reliable foundation for the development and application of uncertainty modeling. 

\subsubsection{Uncertainty quantification}
In AI-empowered software systems, uncertainties present diversity, covering inherent system complexity, data limitations (e.g., noise, sparsity), dynamic operating environment, etc. Especially, epistemic uncertainty, rooted in incomplete knowledge, subjective judgment and assumptions and insufficient data, increases along with the growth of the system complexity. This type of uncertainties is inherently dynamic and complex, which makes it difficult to be described using fixed probability models. In such cases, Bayesian methods quantify uncertainty by utilizing prior knowledge, updating prior probabilities with new observational data, and generating more accurate posterior probability distributions. These methods have been widely applied in software engineering, such as in model-based testing~\cite{camilli2020model} and performance evaluation of configurable software systems~\cite{dorn2020mastering}. Monte Carlo simulations use random sampling to generate large numbers of samples to estimate uncertainty distributions. However, when epistemic uncertainty is high, constructing precise probabilistic models becomes challenging, and this method requires substantial computational resources. Information theory methods utilize entropy, mutual information, and relative entropy to quantify the uncertainty and information content of information or data, thus assessing epistemic uncertainty~\cite{weiss2023uncertainty}. 
Fuzzy logic uses fuzzy sets and logical operators to describe the fuzzy attributes and states of things, as well as the fuzzy relationships between them. It plays an important role in modeling and analyzing epistemic uncertainty and has been widely applied in fields such as expert systems, pattern recognition, and decision support systems~\cite{diaz2020role}.


Practical applications of AI and machine learning face prediction uncertainties. 
Meronen systematically studied uncertainty quantification in deep learning models~\cite{meronen2023uncertainty}, and proposed a Gaussian process-based uncertainty quantification method suitable for resource-constrained environments. To reduce the storage overhead associated with maintaining Markov Chain Monte Carlo (MCMC) samples and demonstrate the practicality of MCMC methods in modern Bayesian neural network applications, Wang et al.~\cite{wang2018adversarial} used Generative Adversarial Networks (GANs) to simulate MCMC samples and obtain a parameterized approximation of their distribution, thus achieving uncertainty quantification. Srivastava et al.~\cite{srivastava2014dropout} used Monte Carlo Dropout as a regularization term for uncertainty computation to avoid posterior probability calculations. Additionally, methods such as deep Gaussian processes~\cite{xu2021accurate} and ensemble-based uncertainty quantification have been applied to quantify prediction uncertainties~\cite{liu2019accurate}.

In addition, with the rapid development and widespread application of machine learning and AI technologies, several open-source uncertainty quantification tools have been released, such as Uncertainty Toolbox~\cite{tran2020methods}, Uncertainty Wizard~\cite{Weiss2021UncertaintyWizard}, and TensorFlow Probability.\footnote{TensorFlow Probability: https://www.tensorflow.org/probability} 
These tools offer a wealth of algorithms for predicting uncertainty and play an important role in uncertainty quantification, calibration, and visualization.

\subsubsection{Uncertainty-aware testing and optimization}
Uncertainty-aware testing explicitly acknowledges uncertainties arising from incomplete requirements, complex system interactions, unpredictable user behaviors, probabilistic outputs of AI-driven components, and dynamic operating environment, when testing software systems.
At the earlier stage of uncertainty-aware testing, Walkinshaw et al.~\cite{walkinshaw2017uncertainty} proposed a black-box testing framework that uses genetic programming to infer the behavioral model of the system under test and selects the most uncertain test cases based on this model. Zhang et al. proposed UncerTest~\cite{zhang2019uncertainty}, which is a model-based testing approach that automatically generates test cases to test CPS under environmental uncertainties. Zhang et al. also proposed UncerPrio~\cite{zhang2023uncertainty} to prioritize high-uncertain test cases for regression testing. For testing self-healing CPS, Ma et al.~\cite{ma2019testing} defined a metric named fragility, based on which an reinforcement learning based testing approach was proposed to test CPS under uncertainties~\cite{ma2019modeling}. 

To automatically generate test oracles for testing the Simulink model of CPS, Menghi et al.~\cite{menghi2019generating} used white noise signals to simulate input uncertainties and uncertain parameter values to simulate unknown hardware choices. Haq et al.~\cite{haq2022efficient} proposed SAMOTA for generating test sets that consider uncertainty in the predictions of alternative models. Shin et al.~\cite{shin2021uncertainty} proposed a UML-based domain-specific language, HITECS, for Hardware-in-Loop (HiL) test case specification and uncertainty-aware analysis methods to check the well-behavedness of HiL test cases. 
Camilli et al.~\cite{camilli2018online} proposed a method that dynamically generates test cases using an uncertainty sampling strategy to deal with uncertain parameters in the Markov decision process model of the Tele Assistance System. Camilli et al.~\cite{camilli2021uncertainty} also leveraged Markov Decision Processes with beliefs attached to transition probabilities, to enable automated generation and selection of test cases. 


\begin{results}[Summary]
Uncertainty characterization in software engineering have matured with an established metamodel and standard. While various quantification methods exist, there is a notable lack of guidelines for selecting appropriate techniques for specific uncertainty types. 
Current uncertainty-aware testing approaches predominantly emphasize external uncertainties (e.g., environmental variability) through black-box testing. However, the literature largely neglects systematic methods to address internal uncertainties and their interactions such as component inter-dependencies. Moreover, the literature has limited exploration of modern architectures like microservices, where distributed workflows and service orchestration introduce unique challenges. 
\end{results}

\section{The Way Forward}\label{future}
After acknowledging the three challenges (Section~\ref{challenges}), in this section we present our vision on establishing an uncertainty-driven system-level testing framework for industrial microservices systems, spanning from unified uncertainty modeling, online detection, causal inference, the impact analysis of uncertainty propagation on system quality in microservices, and all the way to continuous uncertainty-driven fuzzing of microservices. 
\revision{The framework is conceptual and has not yet been fully implemented. However, in this paper we aim to identify key components and workflows that are essential for guiding future development and evaluation.}

\subsection{Architecture}
\begin{figure}[h]
    \centering
    \includegraphics[width=\linewidth]{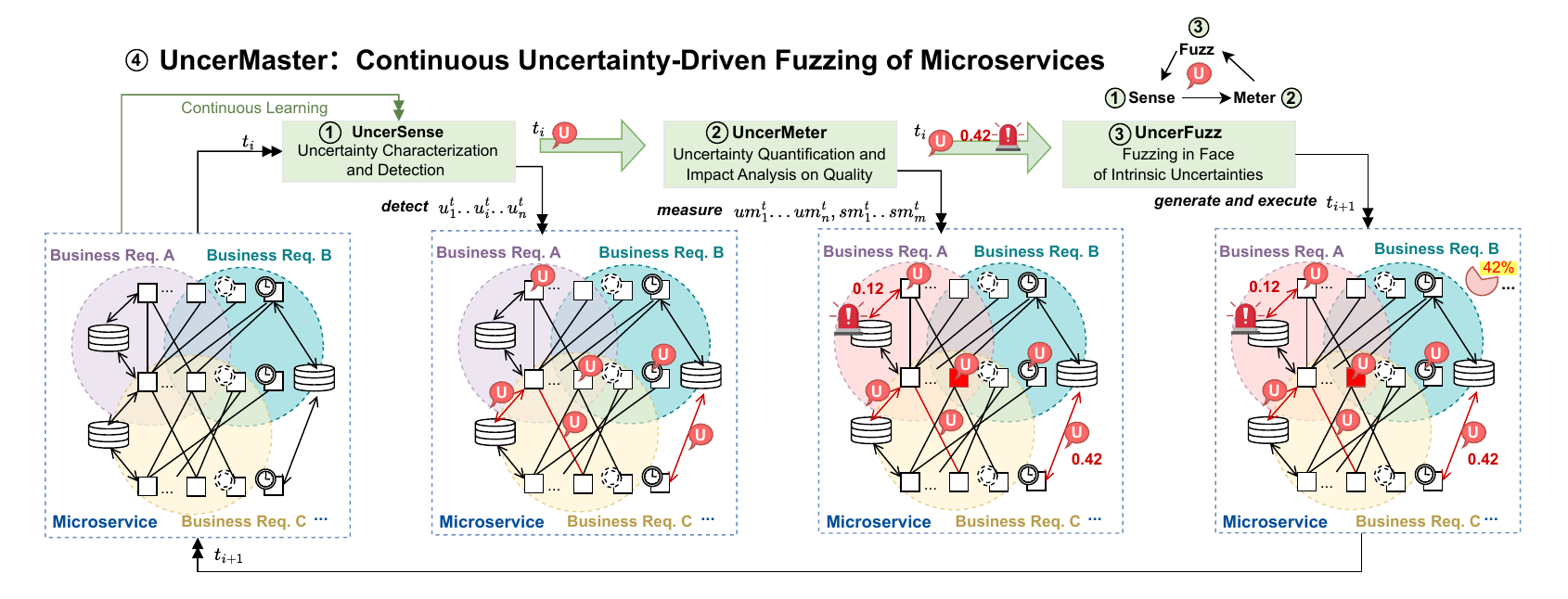}
    \caption{Overview of Continuous Uncertain-driven Fuzzing of Microservices (\uncermaster)}
 \label{fig:overview}
\end{figure}

As shown in Figure~\ref{fig:overview}, the overall architecture of our envisioned uncertainty-driven system-level fuzzing framework for industrial microservices 
\revision{can} advance the state of art through the following four components:
\begin{itemize}
    \item \textit{\uncersense} is the component for online detection of microservice uncertainties, based on architectural characteristics of microservices and industrial production data. The framework will be equipped with an uncertainty detection mechanism which dynamically captures uncertainty occurrences during test execution (e.g., real-time detection of uncertainties $u_1^t, \ldots, u_i^t, \ldots, u_n^t$, where $n$ is the total number of uncertainty occurrences detected by executing test case $t$).  
    \item \textit{\uncermeter} is the component for quantifying uncertainties and their impact on system quality properties. \uncermeter quantifies known uncertainties with fine-grained metrics (e.g., $um_i^t$ denoting the measured uncertainty degree of $u_i^t$), and assess cascading effects on system quality attributes (e.g., $sm_1^t \ldots sm_m^t$ denoting the values of system quality metrics observed when $u_1^t, \ldots, u_i^t, \ldots, u_n^t$ occur). 
    \item \textit{\uncerfuzz} serves as the execution engine for uncertainty-driven microservice testing. It employs a hierarchical, multi- or many-objective optimization algorithm that integrates identified uncertainty occurrences, quantified impacts, and business priorities to generate test cases in favor of high-risk scenarios (e.g., core service call chains) and critical paths. 
    \item \textit{\uncermaster} integrates the above three components and delivers an end-to-end solution for continuous uncertainty detection, uncertainty and impact quantification, and fuzzing.
\end{itemize}

The rest of the section discusses each component and \uncermaster in detail. 

\subsection{Uncertainty Characterization and Detection (\uncersense)}\label{UncerSense}

\subsubsection{Uncertainty classification and characterization}\label{UncerMML}
A comprehensive understanding of uncertainties inherent in microservices including their generative mechanisms (e.g., non-deterministic API behaviors), characteristics (e.g., temporal variability), and taxonomies (e.g., input-, propagation-, and environment-induced uncertainties) forms the foundation to develop uncertainty-aware testing methodologies tailored to microservices. To address this gap, we envision a modeling framework, named Uncertainty-aware Microservices Modeling Language (UncerMML). As recommended practices, UncerMML should be built upon international standards (e.g., PSUM, SysML v2\footnote{SysML v2: https://www.omg.org/spec/SysML/2.0/Beta2/About-SysML}) and stand on the shoulders of existing literature and best practices (e.g., U-Model~\cite{zhang2016understanding} and \evo supporting for both black-box and white-box testing~\cite{arcuri2019restful,zhang2021adaptive,zhang2023rpc}), to enable automated detection of multi-dimensional uncertainty sources and their cascading impacts, thereby addressing the critical absence of holistic uncertainty modeling in microservices research.

The first step towards devising UncerMML involves detecting faults from large-scale operation and testing datasets, by systematically applying traffic recording and replay techniques. To achieve this requires industrial collaborations (to get access to large-scale real-world datasets), systematic detecting uncertainties across multiple dimensions by incorporating core characteristics of microservice architectures, such as distributed communication, heterogeneous technology stacks, and dynamic scalability. For instance, at the architectural level, uncertainties can be classified as exogenous (external inputs), interactive (cross-service dependencies), or endogenous (internal system behaviors), and technical implementations may include nondeterministic program behaviors induced by randomized algorithms (e.g., load balancing), message sequence disorders in partitioned queues (e.g., Kafka), and dynamic routing policy drifts in service meshes (e.g., Istio).
Microservice uncertainties can be characterized in a fine-grained manner from four interrelated aspects: Uncertainty Type ($UT$), Source ($US$), Uncertainty Characteristics ($UC_u$), and Microservice Characteristics ($UC_s$). For example, a message ordering anomaly could be categorized as interactive uncertainty ($UT$), originate from stack misconfigurations ($US$), exhibit randomness and dynamism ($UC_u$), and is associated with the complexity of inter-service communication and data consistency requirements in microservice ($UC_s$). Furthermore, UncerMML should be able to systematically generalize such uncertainty types such as distinguishing epistemic uncertainty (e.g., design flaws in data schemas) from aleatory uncertainty (e.g., inherent randomness like network jitter) to reveal the complete uncertainty propagation from root causes to propagation effects.

\subsubsection{Uncertainty detection}\label{uncertainty detection}
Online uncertainty detection in microservices requires the integration of both black-box and white-box techniques to dynamically monitor both behaviors exposed to the external and internal states, thereby addressing complex scenarios arising from diverse uncertainty sources of various natures. Black-box techniques can rapidly detect exogenous uncertainties or explicit interaction issues by monitoring API request-response patterns, interface throughput fluctuations, and anomalous keywords in logs (e.g., timeouts, retries). However, their inability to observe internal states limits the detection of endogenous uncertainties. To overcome this, white-box techniques  (e.g., \evo~\cite{arcuri2024advanced}) can be used to complement black-box methods by instrumenting code to collect fine-grained data (e.g., function call stacks, control flow paths, and variable state changes) to support the detection of endogenous uncertainties. 

For different uncertainty types defined in the taxonomy of UncerMML, we need to devise different detection strategies. For uncertainties with explicit sources (e.g., non-deterministic programs), predefined functions related to non-deterministic operations (e.g., \texttt{Math.random()} or specific stochastic libraries) can be instrumented with probes to dynamically tag invocations of these functions, enabling real-time detection of code-inherent uncertainties. Additionally, leveraging inherent capabilities of microservice technology stacks (e.g., Istio service mesh for real-time monitoring and distributed tracing), predefined detection rules for scenarios like routing anomalies or traffic spikes can be established. This forms a rule-driven uncertainty detection library to enable systematic uncertainty awareness.

For complex uncertainties with dynamic nature and unclear sources, we need to employ an adaptive learning framework, which can leverage historical data accumulated through traffic recording/replay techniques and already identified uncertainties to train an initial uncertainty detection model via supervised learning. When previously-unknown uncertainties detected during traffic replay or online monitoring, the system adopts an active learning mechanism to filter high-value samples. These samples are validated and injected into incremental training datasets. Online learning algorithms can then dynamically adjust model parameters, updating uncertainty detection boundaries and feature weights in real time, achieving data-driven uncertainty awareness.

\subsection{Uncertainty Quantification and Impact Analysis of Uncertainties on System Quality (\uncermeter)}
Regarding uncertainty quantification in microservices, we must address three key challenges: the dynamic propagation of uncertainties, the opacity of causal relationships, and the complexity of assessing system-wide impact of uncertainties on system quality. An effective approach should integrate white-box testing methodologies, established causal modeling theories and techniques, and a range of uncertainty quantification methods. Specifically, it should support the construction of causal inference models and a multi-dimensional framework for comprehensive assessment.

\subsubsection{Causal inference for uncertainty propagation}\label{causual graphs}
Based on domain knowledge of microservices, we need to systematically analyze dependencies within microservice architectures and exact potential uncertainty propagation paths with code instrumentation, monitoring data, and event logs. Key dependency types in microservices include:   
\begin{itemize}
    \item Explicit invocations: parent-child service dependencies via RPC requests;
    \item Implicit dependencies: asynchronous coupling through Kafka message generation and consumption.
    \item Data dependencies: service-level data contention caused by shared database tables.
\end{itemize}

Specifically, dependencies formed through explicit invocations can be modeled using probabilistic causal graphs, which quantify the conditional dependency strength between services. 
Dealing with implicit dependencies require the integration of temporal event alignment (e.g., matching time windows for message generation and consumption) to construct temporal causal networks. For asynchronous dependencies such as message queues, transfer entropy (TE) can be employed to quantify the information transfer strength between event sequences of services and identify the direction of causality in asynchronous dependencies (e.g., the one-way influence from producer services to consumer services), such that the magnitude of the TE value directly reflects the strength of uncertainty propagation.
For data dependencies, one can leverage data flow provenance techniques to trace propagation paths of shared data states. For instance, for dependencies involving shared resources like databases, a data state transition matrix can be constructed, based on which critical data flow paths can be extracted through techniques such as matrix decomposition, enabling the identification of critical propagation paths. For large-scale industrial services with massive datasets, we need to develop methods for tracing and identifying uncertainty propagation in data dependencies by sampling a small subset of data items. 

All in all, dynamic causal graphs can be synthesized to infer dominant uncertainty propagation paths and critical services, such that systematic characterization of uncertainty propagation can be achieved.

\subsubsection{Uncertainty quantification} \label{UQ}
Existing uncertainty quantification methods predominantly rely on mathematical theories and reasoning frameworks such as probability theory, fuzzy set theory, Bayesian networks, entropy theory, and Dempster-Shafer (D-S) evidence theory. However, uncertainty quantification involves multiple factors, including prior data constraints, types, characteristics and sources of uncertainties, which should be specified in UncerMML (Section~\ref{UncerMML}).

To quantify code-level endogenous uncertainties, we need to understand the sources and behavior of non-deterministic elements in the code, e.g., map random functions to their corresponding probability distribution models. For instance, Java’s standard library functions like \texttt{Math.random()} follow a uniform distribution, while \texttt{Random.nextGaussian()} generates a normal distribution. Third-party libraries (e.g., Apache Commons Math) extend support to complex distributions such as Poisson and exponential distributions.
To quantify these uncertainties, code analysis and runtime instrumentation are required to explicitly identify the distribution type and invocation context (e.g., caller methods, parameter conditions).
At the service-component level, statistical models can be used to capture conditional dependencies among abnormal events—such as service invocation failures, message disorder, slow SQL queries, and resource contention. To quantify the uncertainty associated with individual services, entropy-based metrics offer a systematic measure of unpredictability and variability in their behavior.
For exogenous data uncertainties originating from sensor noise or missing inputs, fuzzy set theory can be applied to quantify membership degrees and confidence intervals, such as defining triangular membership functions.

\subsubsection{Impact analysis of uncertainties on system quality}
Based on the constructed dynamic causal graphs (recall Section~\ref{causual graphs}), uncertainties and their propagation paths can be linked to system quality properties such as availability, throughput, and error rate. For example, sensitivity propagation models (e.g., Monte Carlo simulations) can be employed to quantify the cascading amplification effects of localized faults. By simulating scenarios like the propagation of slow database queries through service call chains, and incorporating factors such as service dependency topology and resource contention probabilities, it becomes possible to predict the impact of these faults on throughput degradation or increases in API timeout rates. In addition, historical traffic data can be used to derive quality degradation correlation functions that map operational uncertainties (e.g., message disorder rates) to business-critical quality properties, such as order fulfillment success rates.

We also need to define a multi-dimensional impact assessment index that combines factors such as business criticality weights (e.g., higher weights for payment services, lower for logging services), real-time quality metrics (e.g., latency and error rates), and causal contribution indicators (e.g., the criticality of propagation paths). 
For instance, index can be calculated by summing up the quality degradation deltas of each microservice weighted by a factor reflecting the centrality of the microservice in the causal graph, i.e., the topological prominence within the service dependency network and its criticality in uncertainty propagation. 

\subsection{Uncertainty-aware Fuzzing of Microservices (\uncerfuzz)}
\begin{figure}[h]
    \centering
    \includegraphics[width=.99\linewidth]{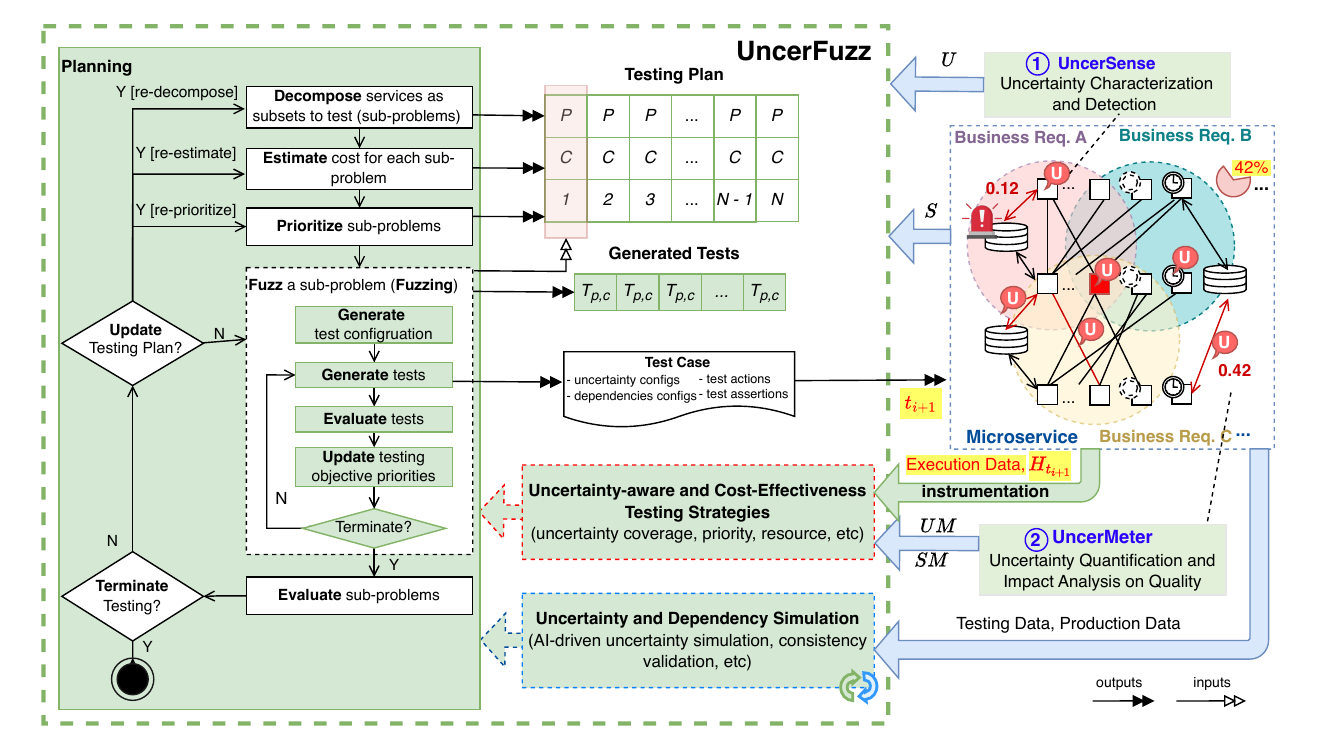}
    \caption{\revision{Overview of Uncertainty-aware Fuzzing of Microservices (\uncerfuzz) and its interactions with \uncersense and \uncermeter}}
 \label{fig:UncerFuzz}
\end{figure}

To be uncertainty-aware, during fuzzing, uncertainties must be actively simulated to expose hidden risks and defects, and to evaluate the dependencies and resilience of microservices under such uncertain conditions. 
In addition, uncertainties must be incorporated into the objectives that guide the fuzzing process. 
Given the inherent complexity of microservices in the presence of uncertainties, decomposition strategies are essential to break down the testing problem (such as a set of services supporting multiple business requirements) into manageable sub-problems (subsets of services to test), enabling an adaptive fuzzing plan tailored for system-level testing of microservices.
Moreover, dynamic fuzzing must be employed to adapt to runtime feedback, evolve testing strategies, and shift priorities of testing objectives, thereby improve the efficiency and effectiveness of fuzzing for each sub-problem.
Hence, an uncertainty-aware fuzzing framework for microservices requires to simulate microservice uncertainties intelligently, explicitly integrate uncertainty information into objectives, and adaptively plan and perform fuzzing of microservices, as illustrated in Figure~\ref{fig:UncerFuzz}.

\subsubsection{Microservice uncertainty simulation}
To simulate microservice uncertainties, it is essential to address the practical limitations: the scale of service deployments and the high resource consumption involved (e.g., clearing an interactive database in a test environment taking over five seconds). To overcome these two challenges, service virtualization technologies can be employed to simulate dependent services or databases, enabling efficient testing without fully deploying the entire system and significantly reducing runtime costs.

To enable intelligent system-level testing of microservices, it is necessary not only to identify and simulate service dependencies but also to inject various types of microservice uncertainties during the testing process. Moreover, it is essential to ensure the consistency of simulated multi-dependency behaviors with real-world logic. More specifically, we need to implement the following components:
\begin{itemize}
    \item \textbf{Identification of multi-source dependencies and enabling simulation for the dependencies.} 1) Based on UncerMML and domain knowledge, automatically identify technology stacks related to microservice dependencies, e.g., JDBC connections, message queue communications, and service invocation chains; 2) For each type of dependency, develop simulation techniques and integrate them to the simulation framework. For instance, for JDBC connections, bytecode instrumentation can be used to capture SQL execution statements and templates, which are then used to generate virtual database schemas and query response logic; for asynchronous messaging middleware (e.g., Kafka, RabbitMQ), message publishing can be simulated, partitioning strategies and the injection of consumption delays can be supported as well; 3) Major protocols can be simulated, such as gRPC via service stub generation, GraphQL with virtual response generation, and REST with virtual HTTP response generation from service stubs.
    \item \textbf{AI-driven simulation of uncertainties and dependencies.} Heuristics-based and data-driven approaches can be employed to enable intelligent simulation of uncertainties and dependencies in microservices. In API fuzzing, the community has automated the generation of mock objects to handle dependencies of external web services via HTTP~\cite{seran2025handling}. These mock responses are guided by white-box heuristics, such as code coverage. However, in the context of uncertainty-aware microservice fuzzing, both uncertainty and a broad range of dependency types remain to be explored.
    Besides heuristics-based approaches, (knowledge enhanced) data-driven solutions can also be employed to create or train agents for simulating uncertainties and dependencies in microservices.
    Using historical traffic recordings (e.g., production data, testing data), multi-modal agents can be trained to handle various simulation tasks, for instance, 1) training sequence-to-sequence models on historical database query results to synthesize mock databases that align with business semantics, 2) employing conditional GANs) 
    to generate responses based on request content that conform to business rules for service behavior simulation, and 3) utilizing reinforcement learning to dynamically adjust the mean and variance of response delays to simulate uncertainties, such as network jitter or service overload.
    \item \textbf{Consistency validation and resolution.} To ensure the quality of 
    uncertainty and dependency simulation, 
    constraints need to be defined and corresponding solvers need to be implemented for checking various types of consistencies such as data consistency and behavioral contract consistency. For example, in an order creation scenario, if service $A$ is simulated to create an order record, the dependent inventory service $B$ must update the simulated database to deduct reserved stock, and subsequent queries must return the updated values accordingly.
\end{itemize}

\subsubsection{Hierarchical, uncertainty-aware fuzzing of microservices}
Compared to single-service testing, system-level testing of microservices faces significantly more complex challenges. For instance, in a scheduling service, the business logic might span over several functional modules, more than 10 core services (excluding dependent services), and near 1000 critical APIs. The testing process further requires participation from external system actors (e.g., corporate employees) to trigger workflows and provide feedback. Such a system-level testing is large-scale, which must simultaneously address massive combinatorial and multi-dimensional efficiency objectives, rendering existing single-service fuzz testing strategies and algorithms inadequate. 

To tackle the dual challenges of combinatorial explosion and resource efficiency in microservice system-level testing, we need a hierarchical, many-objective adaptive testing framework, with the aim to achieve high-efficiency coverage of critical uncertainty scenarios. Details are described below and the key steps and components are also illustrated in Figure~\ref{fig:UncerFuzz}. 
\begin{itemize}
    \item \textbf{Uncertainty-aware testing strategies.} Based on real-time detection of microservice uncertainties, establish a mechanism involving multi-dimensional coverage metrics for test generation and optimization with explicit consideration of uncertainties, including, for instance, \textit{uncertainty type coverage} (e.g., time uncertainty covering the dimensions of the past, current, and future), \textit{uncertainty measurement} (e.g., prioritizing fuzzing on endpoints with high entropy\revision{,} which quantifies the degree of uncertainty in their responses), and \textit{uncertainty coverage} (e.g., proportion of identified uncertainty scenarios exercised). 
    In addition, the strategies incorporate the \textit{priority} and \textit{criticality} of business requirements and associated \textit{risks}, such as dependency path coverage which can be derived from causal graphs to cover uncertainty propagation paths, and  weighted critical service coverage with higher weights assigned to core services based on sensitivity analysis results.
    
    \item \textbf{Test problem decomposition.} Given multi-dimensional indicators (e.g., uncertainty metrics, business priorities, risk levels, and critical paths) and the service dependency topology, employ the divide-and-conquer strategy to partition the testing space into sub-problems, hence transforming a large-scale, complex problem into smaller and more manageable ones.
    \revision{The process can further consider uncertainty correlations and service interaction patterns, which isolate relatively independent uncertainty dimensions and service clusters. Hence, the framework can reduce combinatorial explosion in multi-dimensional uncertainty simulation and avoid to generate redundant or highly overlapping sub-problems.
   Consequently, such structure-aware decomposition enables more efficient allocation of testing resources and facilitates the incremental and scalable exploration of the system-level testing space.}
    
    \item \textbf{Test cost estimation.} Using dynamic causal graphs and Monte Carlo simulation, estimate the testing cost of each sub-problem. The estimated cost is then used to allocate the fuzzing budget.
    \revision{The cost estimation can incorporate multiple factors, including uncertainty dimensionality, service interaction complexity, historical execution time, and resource consumption.
   	Online feedback from ongoing fuzzing steps
    can be used to continuously refine cost estimates and correct inaccurate predictions.
    Such cost estimation can enable informed budget control and help prevent excessive resource consumption during the continued exploration of complex testing scenarios.
    }
    \item \textbf{Test problem prioritization.} Based on the predicted costs, design a scheduling strategy to optimize the order in which sub-problems are solved. 
    Doing so allows for prioritizing simpler and more critical problems, followed by progressively harder or less critical ones. 
    \revision{In addition, prioritization can be combined with adaptive sampling and coverage-guided pruning to optimize the exploration of uncertainty configurations.
   Moreover, by prioritizing computational resources on high-risk and high-impact scenarios, the framework can mitigate high resource demanding required for multi-dimensional uncertainty simulation and combinatorial API interactions, which consequently enables scalable and cost-effective system-level fuzzing in practice.}
    
    \item \textbf{Sub-problem solving (Fuzzing).} Design algorithms to efficiently solve manageable and multi-service fuzzing problems (MSFuzz), along with a many-objective evaluation framework. Such algorithms should integrate intelligent \textit{uncertainty simulation} and \textit{uncertainty-aware testing strategies}. The key steps of the fuzzing is illustrated in Figure~\ref{fig:UncerFuzz}. 
    \begin{itemize}
        \item With the given a sub-problem -- a set of services to test (i.e., $P$), a budget (i.e., $C$), identified uncertainties (i.e., $U$), measured uncertainties (i.e., $UM$) and system quality indicators (i.e., $SM$) associated with $P$, the process begins by generating test configurations to set up the testing environment. This includes tasks such as mocking connected services that are not part of the fuzzing target, initializing required databases, and configuring environmental parameters necessary for simulating uncertainty conditions. 
        \item Guided by heuristic algorithms (e.g., the Many Independent Objective (MIO) algorithm used in \evo~\cite{arcuri2018test,zhang2021adaptive}) or machine learning techniques (e.g., reinforcement learning), a new test case including configurations of uncertainties and dependencies (see \textit{Test Case} in Figure~\ref{fig:UncerFuzz}) can be generated during each run. 
        \item The test is then executed against the services to test ($P$) in microservices. By integrating white-box techniques, runtime information, such as executed paths, performed interactions with external or internal services, achieved new code coverage, and executed nondeterministic programs, can be collected and used to evaluate the test in terms of testing objectives (see \textit{uncertainty--aware testing strategies} in Figure~\ref{fig:UncerFuzz}).
        \item Based on evaluation results, priorities of testing objectives can be updated dynamically, allowing the fuzzing process to adaptively guide future test generation toward high-impact areas affected by uncertainty. 
    \end{itemize}
    
    \item \textbf{Test planning evaluation.} Evaluate the degree of resolution (e.g., solvability) and trends observed when solving each sub-problem, results of which can then be used to dynamically optimize the overall test plan
    \revision{, such as by splitting, merging, or pruning sub-problems.
    The evaluation could also analyze convergence speed, failure discovery rates, and resource utilization patterns, etc., to assess the cost-effectiveness of different testing plans.
    Specifically, sub-problems that exhibit diminishing benefits or consume excessive computational resources can be further decomposed, consolidated with related sub-problems, or pruned to avoid inefficient use of computational resources.
    With this feedback-driven evaluation, it can allow adaptive reconfiguration of testing plans and improve cost-efficiency in large-scale uncertainty-driven fuzzing.}
\end{itemize}

\subsection{\uncermaster}
\uncermaster platform is composed of \uncersense, \uncermeter, and \uncerfuzz, providing continuous uncertainty-driven fuzzing of microservices (see Figure~\ref{fig:overview}). 
Specifically, \uncersense continuously monitors the microservices to infer and detect uncertainties along with their sources. 
\uncermeter analyzes and measures these uncertainties, tracking their propagation across services and providing metrics that reflect the impact of uncertainty on overall system quality. 
\uncerfuzz enables cost-effective, uncertainty-aware fuzzing of microservices in the presence of uncertainties by integrating with intelligent uncertainty simulation, uncertainty-aware testing objectives, adaptive testing plans, and dynamic fuzzing strategies.
Additionally, \uncerfuzz can generate new data that feeds back into \uncersense and \uncermeter, enhancing their ability to sense and measure.
Together, these components enable \uncermaster to establish a continuous cycle of sensing, measuring, and testing, ensuring adaptive and effective fuzzing that evolves with the microservices.

As a good practice, the envisioned platform should adopt a modular architecture, which decouples core algorithms (e.g., dynamic causal inference engines, many-objective optimizers) from low-level white-box processing (e.g., bytecode instrumentation, traffic recording). The platform should also support flexible extensibility through a plugin-based design. The potential architectural layers are as follows. 
\begin{itemize}
    \item Uncertainty awareness layer: Employ dynamic instrumentation to collect real-time internal states of microservices, and integrate distributed tracing data to construct and dynamically update causal graphs, reflecting real-time uncertainty propagation paths, etc.
    \item Dependency simulation layer: Simulate dependencies within a microservice technology stack, including Kafka message brokers, database connections, and RPC service stubs.
    \item Quantitative evaluation layer: Embed an engine for computing values for the uncertainty impact assessment index, which dynamically generates governance priority lists, based on real-time quality metrics and business-critical weights.
    \item Intelligent testing layer: Integrate multi-phase combinatorial optimization algorithms to enable divide-and-conquer strategies and adaptive fuzzing, and generate simulated datasets via AI techniques and leverage the dependency simulation layer to achieve low-cost, high-fidelity test environments.
\end{itemize}


\revision{
\section{Illustrating \uncermaster}\label{sec:example}
We use an e-commerce microservice system (see Figure~\ref{fig:runningexample} for its overview) as an example to illustrate how \uncermaster operates. 
Note that the example is intentionally simplified to illustrate how uncertainty could be represented, detected, and acted upon; hence, it is not meant to be functionally complete or semantically correct. 

The e-commerce system comprises five services as shown in Figure~\ref{fig:runningexample}: \checkout, \payment, \fraud, \order, and \shipping. 
The user-facing entry points are REST APIs in \checkout and \payment. 
Internal service-to-service calls are implemented with gRPC (i.e., \payment $\rightarrow$ \fraud and \order $\rightarrow$ \shipping). 
Cross-service coordination is event-driven via Kafka topics (i.e., \texttt{checkout-topic}, \texttt{payment-topic}, and \texttt{shipment-topic}). 
All services persist states in a shared Postgres database. 
We have open-sourced the running example.\footnote{\url{https://github.com/man-zhang/microservices-projects}}

A typical flow begins when a \textit{User} requests a checkout quote. 
The \checkout service computes the subtotal, tax, and total, and queries an external foreign-exchange (FX) endpoint (Frankfurter) when a currency conversion is needed. 
The quote is persisted as a checkout session and then emitted as a \texttt{checkout-topic} event. 
Next, the \textit{User} submits a payment request to the \payment service, which calls the \fraud service via gRPC to obtain a risk decision and score. 
If approved, the \payment service invokes a payment gateway client to obtain a transaction identifier. 
The payment result is persisted and published as a \texttt{payment-topic} event. 
The \order service consumes \texttt{payment-topic} events and creates an order-confirmation record. 
For approved payments, \order calls the \shipping service via gRPC to create a shipment. 
The \shipping service persists the shipment, contacts an external logistics endpoint to dispatch it, and publishes a \texttt{shipment-topic} event. 
This mixed RPC--event chain decouples services while still forming an end-to-end business flow. 

\begin{figure}
	\centering
	\includegraphics[width=0.99\linewidth]{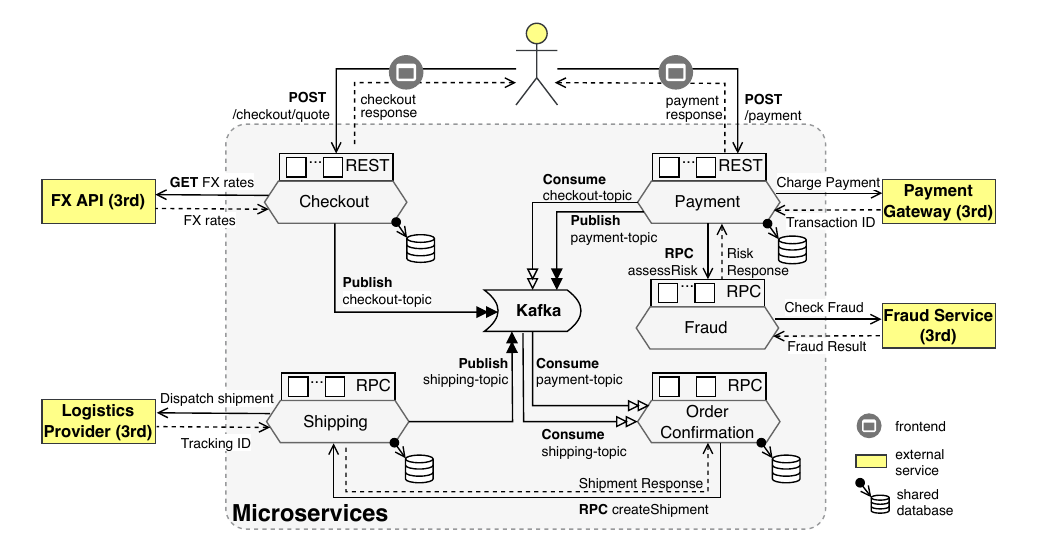}
	\caption{\revision{Microservices and their main interactions in the e-commerce example}}
	\label{fig:runningexample}
\end{figure}

Even in this simplified setting, the system is exposed to heterogeneous sources of uncertainty. 
Some uncertainties are induced by external dependencies (e.g., FX and logistics endpoints) that may change over time or fail transiently. 
Some uncertainties are introduced by threshold-based decisions (e.g., fraud scores close to decision boundaries) and by hidden or evolving states (e.g., history-dependent fraud features). 
Some uncertainties are caused by distributed coordination (e.g., asynchronous consumption and partial progress when an RPC fails after a database write). 
In the following subsections, we illustrate what \uncermaster would observe and how it would exploit such uncertainty during system-level fuzzing. 
}

\revision{
\subsection{Uncertainty Detection with \uncersense}
\uncersense plays the role of monitoring runtime information in order to identify uncertainty occurrences during test execution. 
Such information includes service logs, actuator/prometheus metrics, and interaction metadata across REST, gRPC, Kafka, and database operations. 
Detected events are then mapped to \uncermml attributes and to predefined modeling assumptions for known uncertainty behaviors. 
In this paper, we focus on what information requires to be collected to enable identification, rather than on the concrete uncertainty detection mechanisms, which are left as future work and could be rule-based, data-driven, or even hybrid. 

In the example, uncertainty can be introduced by input processing, decision logic, external dependencies, and asynchronous propagation. 
However, uncertainties manifest differently across services and could follow different statistical distributions.
For instance, monetary totals in \checkout are computed using \texttt{BigDecimal} and deterministically rounded to two decimals, which introduces no stochastic numerical variation by itself.
Variability is actually primarily driven by time-varying external signals, e.g., periodic refreshes of currency-exchange rates from an external provider (denoted as FX rates),\footnote{In this paper, ``FX'' denotes foreign-exchange (currency-exchange) rates. 
For instance, in the running example, \checkout refreshes its cached rates by querying the Frankfurter API (e.g., \url{https://api.frankfurter.app/latest?from=USD&to=EUR}), as documented at \url{https://www.frankfurter.app/docs}.} and by discrete thresholding such as fraud decisions made by Fraud Service provided by the third party. 

For input-processing services, \uncersense collects a lightweight, service-specific record of inputs and computed outputs. 
In \checkout, this includes the request payload (items, quantities, and target currency) and the computed outputs (subtotal, tax, and total). 
To explain variability, \uncersense also records currency-conversion metadata, such as the exchange-rate value used and whether it was served from cache, refreshed from the live endpoint, or obtained via fallback. 
Under the same basket, small differences can arise when cached FX rates are refreshed, and larger discontinuities can arise when the service falls back to predefined rates after a failed refresh. 

For uncertainty related to external dependencies, \uncersense collects dependency identifiers, timestamps, outcomes, and latency. 
In the example, main external HTTP dependencies include the \checkout call to the Frankfurter FX API, the \payment call to a payment gateway, and the \shipping call to a logistics provider. 
Note that, in this example, the payment gateway client in \payment is implemented as a local stub that simply generates a transaction ID. 
However, in real deployments, the external payment gateway is often a major source of uncertainty due to variable latency, transient failures, and provider-side throttling or outages.\footnote{\url{https://stripe.com/en-nl/resources/more/payment-gateway-down}}

For decision-making services, \uncersense captures both the final decision and the intermediate evidence used to reach it. 
In \fraud, the gRPC service computes a score and assigns a discrete decision using fixed thresholds (0.3 and 0.7). 
The scoring logic also produces an explicit confidence value derived from an ensemble-style predictive distribution. 
Accordingly, \uncersense records the request features (e.g., \texttt{userId} and amount), intermediate factors (e.g., history count and external signal), and the resulting score, confidence, and decision. 
Uncertainty is operationalized as increased decision instability: near decision boundaries, small shifts in features or history may flip the output among \texttt{APPROVE}/\texttt{REVIEW}/\texttt{REJECT}. 

For propagated and asynchronous workflows, \uncersense focuses on correlation and state transitions across services. 
In \order, \uncersense records correlation keys (e.g., \texttt{orderId}, \texttt{paymentId}, and \texttt{shipmentId}), producer timestamps from events, consumer receive time, and the service-side state transitions written to the database. 
These records enable identifying partial progress, such as when an order-confirmation record is persisted but the corresponding shipment creation remains unresolved because the gRPC call to \shipping fails. 
In \shipping, \uncersense similarly records dispatch outcomes, response timestamps, and whether a tracking ID is present, since shipments may exist without downstream dispatch completion under failure conditions. 

Across all interaction types (service $\leftrightarrow$ service, service $\leftrightarrow$ Kafka, service $\leftrightarrow$ external HTTP, and service $\leftrightarrow$ database), \uncersense can rely on a unified core schema. 
This schema includes correlation identifiers, timestamps, interaction type, and execution context (service, endpoint/method/topic). 
Because distributed tracing headers are not propagated by default in the reference implementation, business identifiers serve as the primary linkage for cross-service correlation. 
Based on these observations, \uncersense dynamically identifies uncertainty instances $u_1^t, \ldots, u_n^t$ during the execution of test case $t$. 
Each uncertainty instance is associated with its execution context and evidence sources. 
Potential uncertainties considered in the example are summarized in Table~\ref{tab:example} for the illustration purpose and hence it is by no means to be complete. Empirical studies are needed in the future to systematically characterize such uncertainties and develop a comprehensive taxonomy (i.e., \uncermml) for industrial microservice systems.
}

\begin{table}[t] 
	\centering 
	\caption{\revision{Potential sources and types of uncertainty in the e-commerce example}} 
	\label{tab:example} 
	\resizebox{.99\linewidth}{!}{ 
		\revision{
	\begin{tabular}{p{.11\textwidth} p{.18\textwidth} p{.93\textwidth}}
		\toprule
		Service
		& System Context
		& Description \\
		\midrule
		
		Checkout
		& Input Processing; \newline External Dependency
		& Price, tax, and currency conversion are computed with deterministic rounding, while variability is driven by currency-exchange rate refresh, caching, and fallback behavior. 
		\newline \textbf{UT:} Input- and environment-induced uncertainty. 
		\newline \textbf{US (Aleatory):} Time-varying currency-exchange rates and transient failures when calling the Frankfurter API. 
		\newline \textbf{US (Epistemic):} Rate freshness and provenance (cache vs.\ live vs.\ fallback) unless explicitly instrumented. 
		\newline \textbf{UC$_u$:} Temporal variability in quoted totals under identical baskets. 
		\newline \textbf{UC$_s$:} Currency-exchange client, cache TTL, scheduled refresh with retries, and fallback-rate path. 
		\newline \textbf{Modeling/Measurement:} $x$ as a distribution over observed totals across time and refresh regimes. 
		\newline \textbf{Example $um$ (hypothetical):} $um_1^t=P(\text{rateSource}=\text{fallback})=0.12$, $um_2^t=p95(L_{\text{fx}})=1.8\text{s}$. 
		\newline \textbf{Quantification Tools:} Actuator/Prometheus metrics and logs, TensorFlow Probability for distribution fitting (e.g., tail latency). \\
		
		Payment
		& Communication; \newline Service Dependency
		& Payment processing depends on a gRPC call to \fraud to obtain a risk decision, and it emits \texttt{payment-topic} events for downstream processing. 
		\newline \textbf{UT:} Interaction- and decision-induced uncertainty. 
		\newline \textbf{US (Aleatory):} Variable gRPC latency and transient failures in service-to-service communication. 
		\newline \textbf{US (Epistemic):} Limited visibility into downstream decision rationale and evolving history-dependent features. 
		\newline \textbf{UC$_u$:} Status instability when risk scores are near decision thresholds and when dependency health fluctuates. 
		\newline \textbf{UC$_s$:} gRPC dependency on \fraud and event publication to Kafka. 
		\newline \textbf{Modeling/Measurement:} $T$ as a distribution over request latency and $S$ as a categorical payment outcome. 
		\newline \textbf{Example $sm$ (hypothetical):} $sm_{\text{review}}^t=P(\text{paymentStatus}=\text{REVIEW})$. 
		\newline \textbf{Quantification Tools:} Actuator/Prometheus metrics and logs, TensorFlow Probability for latency modeling. \\
		
		Fraud
		& AI/ML \newline Inference; \newline External Signal
		& Risk scoring follows an AI/ML inference workflow that outputs a score and an uncertainty indicator (e.g., confidence), which is then mapped to a discrete decision via fixed thresholds. 
		\newline \textbf{UT:} Inference- and decision-induced uncertainty. 
		\newline \textbf{US (Epistemic):} Model approximation, limited/biased training data, and concept drift that reduce fidelity to real fraud dynamics. 
		\newline \textbf{US (Aleatory):} Noisy and time-varying behavioral signals and external assessments in realistic deployments. 
		\newline \textbf{UC$_u$:} Predictive dispersion and decision flips near thresholds (0.3 and 0.7). 
		\newline \textbf{UC$_s$:} Model inference pipeline, feature dependency on historical state, and threshold-based post-processing. 
		\newline \textbf{Modeling/Measurement:} $y$ as a predictive distribution, with uncertainty summarized by variance/entropy and confidence intervals. 
		\newline \textbf{Example $um$ (hypothetical):} $um_3^t=1-\overline{\text{confidence}}=0.27$. 
		\newline \textbf{Quantification Tools:} Uncertainty Wizard for predictive distributions (e.g., MC dropout), and Uncertainty Toolbox for calibration metrics; Actuator/Prometheus metrics and stored fraud-check records support telemetry. \\
		
		Order\newline Confirmation
		& Event-Driven \newline Aggregation; \newline Communication
		& The service consumes \texttt{payment-topic} events, persists an order-confirmation record, and calls \shipping via gRPC for approved payments. 
		\newline \textbf{UT:} Propagation- and partial-progress uncertainty. 
		\newline \textbf{US (Epistemic):} Partial observability of event timing, consumer backlog, and failure points across a multi-step workflow. 
		\newline \textbf{US (Aleatory):} Broker delay, consumer lag, and transient RPC failures to \shipping. 
		\newline \textbf{UC$_u$:} Ambiguous intermediate states when a DB write succeeds but the downstream gRPC call fails to complete. 
		\newline \textbf{UC$_s$:} At-least-once event delivery, idempotence checks, and eventual consistency across DB and RPC effects. 
		\newline \textbf{Modeling/Measurement:} $S$ as a categorical distribution over confirmation states and $D$ as a distribution over end-to-end delays. 
		\newline \textbf{Example $sm$ (hypothetical):} $sm_{\text{stuck}}^t=P(\text{approvedPayment}\wedge \text{shipmentId}=\varnothing)$. 
		\newline \textbf{Quantification Tools:} Actuator/Prometheus metrics, application logs, and database state snapshots. \\
		
		Shipping
		& Communication; \newline External Dependency
		& Shipment creation is triggered via gRPC, persisted immediately, and followed by an external logistics dispatch call that may be slow or fail. 
		\newline \textbf{UT:} Environment- and propagation-induced uncertainty. 
		\newline \textbf{US (Aleatory):} Stochastic logistics endpoint latency and availability. 
		\newline \textbf{US (Epistemic):} Limited visibility into external carrier state and dispatch completion. 
		\newline \textbf{UC$_u$:} Missing-\texttt{trackingId} outcomes and delayed dispatch completion under failure. 
		\newline \textbf{UC$_s$:} Persist-then-dispatch workflow and external HTTP integration. 
		\newline \textbf{Modeling/Measurement:} $T$ as a long-tail distribution over dispatch latency and $M$ as a missingness indicator for tracking IDs. 
		\newline \textbf{Example $um$/$sm$ (hypothetical):} $um_4^t=p95(L_{\text{logistics}})=3.5\text{s}$, $sm_{\text{missingTrack}}^t=P(\text{trackingId}=\varnothing)$. 
		\newline \textbf{Quantification Tools:} Actuator/Prometheus metrics, logs, database state inspection, and TensorFlow Probability for latency modeling. \\
		
		\midrule
		Cross-Service
		& Impact \newline Analysis
		& \textbf{Impact model (hypothetical):} $sm_{\text{missingTrack}}^t \approx \beta_1\,p95(L_{\text{logistics}})+\beta_2\,P(\text{rateSource}=\text{fallback})+\beta_3\,(1-\overline{\text{confidence}})$, where $\beta_1,\beta_2,\beta_3$ are fitted from observations and used to prioritize uncertainty-based fuzzing. \\
		
		\bottomrule
	\end{tabular}
} 
	} 
\end{table}

\revision{
\subsection{Uncertainty Quantification with \uncermeter}

After uncertainty instances are detected, \uncermeter quantifies their magnitude and estimates their impact on system-level quality attributes. 
Once again, all modeling choices below are illustrative and are presented to demonstrate how the framework could operationalize uncertainty in a test loop. 

For numerical variation in \checkout, \uncermeter can model observed totals as a random variable and estimate distribution parameters from runtime samples. 
For example, it may fit a Gaussian model and compute $\mu_x$ and $\sigma_x$ from repeated executions under comparable conditions. 
A non-trivial $\sigma_x$ then indicates that rate refresh, fallback, and rounding effects can measurably perturb totals. 
For interaction-induced uncertainty (e.g., external calls and RPCs), \uncermeter can model latency using heavy-tailed distributions such as log-normal. 
Percentile-based estimators can be used to derive $\mu_T$ and $\sigma_T$ from observed response-time samples. 
Long-tail latency, such as a 99th percentile exceeding a timeout threshold, can then be related to elevated retries and partial execution. 
For decision uncertainty in \fraud, \uncermeter can treat prediction outputs as stochastic variables around an expected score. 
The model-provided confidence and dispersion signals can be used to quantify instability near decision boundaries. 
These measurements can be aggregated per user cohort, amount range, or history regime to support targeted testing. 
For propagation uncertainty in \order, \uncermeter can estimate categorical state probabilities of order-confirmation outcomes. 
A Bayesian model can infer the probability mass of states such as \texttt{CREATED}, \texttt{FAILED}, and \texttt{SHIPMENT\_REQUESTED} under different injected conditions. 


%
%

Regarding uncertainty quantification tools, \uncermeter could integrate existing tools (such as Uncertainty Wizard~\cite{Weiss2021UncertaintyWizard}, TensorFlow Probability\footnote{TensorFlow: \url{https://www.tensorflow.org/probability}} and Uncertainty Toolbox~\cite{tran2020methods}) and libraries for uncertainty modeling, measurement, and diagnostics, depending on component type and available telemetry.
Table~\ref{tab:example} represents the potential applicability of modeling approaches and corresponding tools for quantifying uncertainties.
For example, the \fraud service employs a neural classifier to output a risk score $y\in[0,1]$. 
With Uncertainty Wizard, Monte Carlo dropout is enabled and $K$ stochastic forward passes are executed for the same request, which yields: $\{y^{(k)}\}_{k=1}^{K}$. 
The predictive mean $\mu=\frac{1}{K}\sum_{k} y^{(k)}$ and variance $\sigma^2=\frac{1}{K}\sum_{k}(y^{(k)}-\mu)^2$ can be then computed to quantify epistemic uncertainty induced by model stochasticity. 
We can further compute Bernoulli predictive entropy $H(\mu)=-\mu\log\mu-(1-\mu)\log(1-\mu)$ by using the predictive mean $\mu$ to
capture overall predictive uncertainty and decision ambiguity near classification thresholds.  
This yields $um_3^t=1-\overline{\text{confidence}}$ in Table~\ref{tab:example}.
For the external logistics dispatch call in \shipping, latency $L$ is modeled as a log-normal random variable. 
TensorFlow Probability fits parameters $(\mu_T,\sigma_T)$ from runtime samples and computes tail metrics such as $p95(L)$, which are used in $um_4^t$. 
This approach is suitable when uncertainty manifests as variability in continuous observations and can be adequately characterized by a known distribution family.
For predictive components such as fraud risk models, Uncertainty Toolbox can be used to compute calibration metrics including expected calibration error (ECE), negative log-likelihood (NLL), and Brier score. 
These diagnostics quantify the reliability of reported confidence values and support threshold selection for uncertainty-driven fuzzing.

Regarding system-level quality metrics, during continuous testing and fuzzing, \uncermeter aggregates execution observations into uncertainty metrics $um^t$ and system-level outcomes $sm^t$ for each test case $t$. 
As summarized in Table~\ref{tab:example}, uncertainty metrics can capture service-local phenomena such as fallback-rate usage and external-call tail latency in \checkout, inference uncertainty in \fraud, and logistics-dispatch tail latency in \shipping. 
Table~\ref{tab:example} also illustrates system-level outcome indicators, including the fraction of payments routed to \texttt{REVIEW}, the fraction of approved payments whose order confirmations remain stuck without shipment association, and the fraction of shipments created without tracking identifiers. 
Given these measurements, \uncermeter can fit impact models that relate $um^t$ to downstream quality outcomes in $sm^t$. 
Table~\ref{tab:example} provides an illustrative model where missing tracking identifiers are explained by logistics tail latency, fallback-rate usage, and fraud-model uncertainty, with coefficients fitted from observed executions. 
These learned relationships are then used by \uncermaster to prioritize uncertainty-driven fuzzing toward the services and interaction paths with the highest inferred impact.

While the above examples illustrate the feasibility of leveraging existing uncertainty quantification approaches or tools, 
determining which one to use, how to configure it, and how to interpret its outputs in any real-world microservice system remains a non-trivial challenge. 
The suitability of a given technique depends on factors such as workload characteristics, data availability, latency constraints, and cross-service dependencies. 
Systematic guidelines for selecting and integrating uncertainty quantification tools in large-scale production systems therefore require further investigation and constitute an important direction for future work.

%
%
}

\revision{
\subsection{Uncertainty-Driven Fuzzing with \uncerfuzz}

Based on the uncertainty measurements and impact assessments produced by \uncermeter, \uncerfuzz aims to generate test cases that prioritize uncertainty-sensitive behaviors and high-impact scenarios. 
The goal is not to maximize coverage randomly,
but to guide exploration toward executions where uncertainty is likely to propagate and expose potential system issues. 

In each test cycle, \uncerfuzz consumes detected uncertainty instances and their contexts from \uncersense, and quantified uncertainty metrics $um_i^t$ and system-level indicators $sm_j^t$ from \uncermeter. 
It then constructs uncertainty-aware testing objectives and an adaptive testing plan that reflects both uncertainty severity and cost constraints. 
Because budgets are limited, \uncerfuzz may only actively fuzz a subset of services or call chains at a time, while keeping the remaining parts fixed. 
Within the selected scope, \uncerfuzz synthesizes tests using dynamic fuzzing strategies and intelligent uncertainty simulation. 
For computation-centric uncertainty in \checkout, test generation can emphasize currency selections and execution timing that stress rate caching and fallback behavior. 
For decision-centric uncertainty in \fraud, tests can emphasize inputs that drive inference near decision boundaries, where small perturbations can flip outcomes. 
For propagation-centric uncertainty in \order and \shipping, tests can emphasize partial-progress executions where persistence succeeds but downstream gRPC calls or external dispatch calls do not complete. 


Consider a fuzzing iteration in which \uncerfuzz determines the campaign budget to a single service, e.g., \fraud, because \uncermeter observes increased instability of inference outcomes near the decision thresholds.  
In this iteration, \uncerfuzz focuses only on the \payment $\rightarrow$ \fraud gRPC interaction and treats the payment gateway and downstream order/fulfillment services as out of scope. 
It constructs a test objective that maximizes decision sensitivity around the thresholds, e.g., maximizing the fraction of executions whose predicted score falls within a narrow band around 0.3 or 0.7 while also maximizing the observed variance of scores and minimizing confidence. 
To implement this objective, \uncerfuzz generates \texttt{RiskRequest} inputs that systematically vary amount and user history conditions, aiming to drive the inferred score into the threshold-adjacent regions where small perturbations can flip \texttt{APPROVE}/\texttt{REVIEW}/\texttt{REJECT}. 
The test flags a fault indicator when minor input perturbations cause large swings in decision outcomes or when confidence drops sharply without corresponding explanatory changes in input features, which indicates brittle inference behavior that may amplify downstream uncertainty. 
The resulting traces and quantified metrics are fed back into \uncersense and \uncermeter, enabling \uncerfuzz to decide whether to broaden the campaign to downstream services or to refine the inference-focused fuzzing objective. 
}

\revision{
\subsection{Iterative and Continuous Fuzzing with \uncermaster}
The \uncermaster platform establishes a continuous cycle of sensing, measuring, and testing by integrating \uncersense, \uncermeter, and \uncerfuzz. 
It orchestrates the end-to-end workflow, coordinates data exchange among the components, and manages iterative testing cycles. 
During the cycle, as a new uncertainty is observed and quantified, \uncermaster adaptively updates testing objectives, selects where to invest testing budget, and steers fuzzing strategies accordingly. 
\uncermaster aims to detect uncertainty-induced fault trends before they are triggered at scale by end-user traffic. 
In our vision, this is achieved by continuously extracting uncertainty characteristics from execution info, quantifying their propagation and impact, and launching focused fuzzing campaigns when risk signals rise. 
Below, we illustrate two examples to demonstrate how \uncermaster can expose such faults through iterative uncertainty-driven testing. 

The first example concerns checkout quoting under an external currency-exchange dependency. 
Assume that \uncersense observes intermittent timeouts or errors when \checkout queries the external currency-exchange endpoint and that quote totals exhibit step changes for identical baskets. 
\uncersense correlates quote outputs with the applied exchange-rate values and their provenance (cache, live fetch, or fallback), and instantiates uncertainty occurrences $u_i^t$ that reflect external dependency uncertainty interacting with deterministic monetary rounding. 
\uncermeter then quantifies quote instability and estimates its impact on system-level outcomes such as price consistency. 
If the estimated risk exceeds a policy threshold, \uncerfuzz prioritizes \checkout and reproduces the step-change pattern using uncertainty simulation (e.g., injecting controlled failures into the currency-exchange endpoint and varying cache settings). 

The second example concerns fulfillment progress along the \order $\rightarrow$ \shipping $\rightarrow$ external logistics chain. 
Assume that \uncersense observes increasing tail latency and intermittent failures for the external logistics dispatch call in \shipping. 
Also assume that \uncersense observes an increasing rate of approved payments for which \order persists an order-confirmation record, but the downstream gRPC call to \shipping does not complete within an expected time window. 
To support diagnosis, \uncersense links evidence across services using business identifiers (e.g., \texttt{orderId} and \texttt{paymentId}) and timestamps extracted from REST requests, Kafka events, and gRPC calls, thereby instantiating uncertainty occurrences $u_i^t$ along the propagation path. 
\uncermeter estimates the distribution of dispatch latency and the probability of partial progress, and derives system-level indicators (e.g., missing shipping progress and missing tracking identifiers). 
If these indicators exceed policy thresholds, \uncerfuzz launches a focused campaign that prioritizes testing towards the \order and \shipping path and steers \uncerfuzz to generate tests to cover the path. 
For example, \uncerfuzz can combine injected logistics uncertainty with bursty payment flows to amplify consumer pressure and timing skew, which increases the likelihood of partial progress and inconsistent cross-service state. 

In both examples, faults do not need to be single deterministic bugs in isolated components, but system-level failure modes that emerge when uncertainty interacts with persistence boundaries, asynchronous processing, and downstream calls. 
The essence of \uncermaster is to treat rising quote instability, partial progress, and inconsistent cross-service states as measurable uncertainty-propagation signals, which are used to proactively construct tests before a large fraction of end-user requests experience the issue. 
This aligns with our vision that uncertainty-aware system-level testing can serve as an early-warning and pre-development validation loop for industrial microservice systems. 

}

\revision{
\section{Discussions}\label{sec:discussions}

\subsection{Bootstrapping and Cold-Start Challenges}

At the beginning, system-level fuzzing may face the ``cold-start'' problem due to the lack of sufficient observations of system-level failures~\cite{zhang2025failure}. 
With \uncermaster, the framework can initially leverage available domain knowledge, system specifications, architectural dependencies, and coarse-grained monitoring data to construct a preliminary uncertainty and causal model using \uncermml, such as uncertainties associated with input processing, AI/ML-based decision making, and external dependencies. 
In addition, the framework can enable a bootstrapping phase in which \uncerfuzz performs exploratory fuzzing and generates diverse executions under different uncertainty perturbations to provide informative inputs for \uncersense and \uncermeter. 
Furthermore, seeding is a commonly adopted approach in industrial settings~\cite{zhang2024seeding}, and the initial causal model (e.g., constructed by domain experts) can serve as a seed to guide early-stage fuzzing and analysis.

As testing proceeds, \uncerfuzz continuously produces new execution traces, uncertainty injection records, and failure observations. 
These runtime data are incrementally fed to \uncersense and \uncermeter, which allows them to identify uncertainty and refine uncertainty measurements and causal inferences. 
Through this iterative feedback loop, the causal model is progressively improved to enable more accurate identification of high-risk behaviors and more effective fuzzing guidance, which eventually may allow the framework to gradually overcome the cold-start limitation.

However, we acknowledge that this iterative refinement process may require substantial time and resources, and its effectiveness may depend on the availability and quality of runtime data. 
Addressing these practical constraints remains an important future research direction.

\subsection{Challenges of Evaluating \uncermaster In Industrial Settings}

Setting up and empirically evaluating a system-level testing framework for microservices in industrial settings poses significant challenges.
Prior industrial experience with microservice fuzzing (e.g., the \evo user studies) highlights that even deploying and evaluating a state-of-the-art fuzzer on production-grade services requires non-trivial engineering effort and close collaboration with practitioners, due to issues such as environment setup, dependency management, and integration into existing development workflows~\cite{zhang2025industry}. 

First, industrial microservices may consist of hundreds or thousands of loosely coupled services implemented in diverse technologies and deployed across rapidly changing infrastructures. System-level fuzzing such systems at a realistic scale often demands testing environments provided by  the industrial partner.
In addition, a major barrier to system-level testing is the prevalence of external and platform dependencies, including payment gateways, authentication services, recommendation engines, inventory services, and various middlewares. 
Running tests at scale without disrupting operations commonly requires controlled substitutes (mocks/stubs)~\cite{seran2025handling,arcuri2020sql} as well as systematic \emph{seeding} of system states (e.g., accounts, coupons, inventories, and feature flags) to reach deeper behaviors. 
An industrial case study~\cite{zhang2024seeding} on enterprise RPC API fuzzing shows that seeding and mocking are not optional enhancements but practical prerequisites for making white-box fuzzing effective and deployable in enterprise environments.

Beyond test execution, collecting high-quality runtime telemetry such as distributed traces, failure logs, and uncertainty-related measurements, often requires deep integration with production monitoring and observability infrastructures~\cite{li2022enjoy}. 
In industrial environments, such integration may be constrained by security policies, privacy regulations, and operational risks~\cite{arcuri2025fuzzing}. 
Moreover, instrumentation itself introduces additional challenges~\cite{hammad2025empirical,nou2025investigating}. 
Although lightweight instrumentation is acceptable in many cases, monitoring overhead and system perturbations may still affect latency-sensitive services, interfere with timing-dependent behaviors, and complicate reproducibility.
These effects can lead to spurious findings (e.g., false positives) or hinder reliable replay of test outcomes, thereby requiring careful engineering and further empirical investigation. 
Thus, the effectiveness of our framework depends on the availability of high-quality traces, logs, and metrics with consistent correlation identifiers across services, as well as principled definitions of system impact (e.g., tail-latency shifts, error-budget consumption, or partial outages). 
Existing studies on microservices highlight the benefits of such instrumentation and the substantial engineering effort required to obtain actionable and attributable evidence~\cite{li2022enjoy}. 
Finally, fuzzing and its evaluation often necessitates long-running studies to capture rare but high-impact failures, which further increases the required time and resource investment~\cite{zhang2025industry,klees2018evaluating}.

These realities motivate a staged development and evaluation strategy:
(1) begin with a representative subset of services and controlled dependencies (via seeding/mocking), 
(2) progressively expand the scope and realism of uncertainty and fault scenarios, and 
(3) collaborate with industry partners to access realistic telemetry and workloads under strict safety controls. 
Our framework is designed with these constraints in mind, aiming to be incrementally deployable and empirically assessable under practical industrial limitations.

}

\section{Concluding Remarks}\label{conclusion}
Microservices have been widely adopted across various domains, with significant effort dedicated to ensuring their dependability through testing. Most existing testing approaches rely on fuzzing techniques, mostly focusing on API fuzzing, which tests individual services. However, industrial microservices often contain many complex, heterogeneous, dynamic, and interrelated uncertainties that propagate across multiple services, which substantially impact overall system quality. These challenges become increasingly prominent and prevalent as microservices systems inevitably are adopting more and more AI components, which introduces extra layers of uncertainties. These uncertainties have rarely been explicitly acknowledged and systematically integrated into system-level testing strategies for microservices. 

In this paper, we argue for the importance and significance of systematically addressing multi-dimensional uncertainties in microservice fuzzing. We propose a concrete vision, named \uncermaster, which is composed of three key components: \uncersense, \uncermeter, and \uncerfuzz, by holistically leveraging various types of technologies (e.g., Generative Adversarial Networks, uncertainty quantification), standards (e.g., System Modeling Language, Precise Semantics for Uncertainty Modeling), and best practices (e.g., white-box fuzzing, service virtualization). \revision{We also developed an e-commerce example and used it to illustrate \uncermaster systematically.}
We hope readers find this vision valuable and join us in advancing efforts to realize this framework.
Currently, we are developing \uncermaster as an open-source platform, initially targeting JVM-based microservices in collaboration with our industrial partners. 
However, the platform is designed to be extensible to other programming languages. 
We are confident that its implementation will benefit the community and contribute to ensuring the dependability and trustworthiness of industrial microservices, using and building on top of state-of-the-art open-source fuzzers such as \evo. 

\section*{Acknowledgements}
This work is supported by the National Science Foundation of China (grant agreement No. 62502022).
Andrea Arcuri is funded by the European Research Council (ERC) under the European Union’s Horizon 2020 research and innovation programme (EAST project, grant agreement No. 864972).


\bibliographystyle{ACM-Reference-Format} 

%




\end{document}